# The Heat Conduction Renaissance

Aditya Sood[1,2], Eric Pop[1,3], Mehdi Asheghi[2], and Kenneth E. Goodson[2,3]
[1]Department of Electrical Engineering, Stanford University, CA 94305, USA
[2]Department of Mechanical Engineering, Stanford University, CA 94305, USA
[3]Department of Materials Science and Engineering, Stanford University, CA 94305, USA
Email: goodson@stanford.edu

**Abstract**

Some of the most exciting recent advancements in heat conduction physics have been motivated, enabled, or achieved by the thermal management community that ITherm serves so effectively. In this paper we highlight the resulting *renaissance* in basic heat conduction research, which is linked to cooling challenges from power transistors to portables. Examples include phonon transport and scattering in nanotransistors, engineered high-conductivity composites, modulated conductivity through phase transitions, as well as the surprising transport properties of low-dimensional (1D and 2D) nanomaterials. This work benefits strongly from decades of collaboration and leadership from the semiconductor industry.

**Introductory Comments**

The hallmark of continued "Moore's law" scaling of electronic chips and the associated increases in power density has been the increasing fidelity of nanofabrication technology. This has yielded electronic nanostructures of breathtaking complexity and near-atomistic precision for microprocessors, memory, and optoelectronics. In parallel, the increasing spatial packing density, from portables to power electronics, has set a high bar for the heat conduction and thermomechanical properties required in electronic packaging [1].

This has brought heat conduction engineering – both in packaging and in semiconductor nanostructures – to the forefront of the design and engineering of most modern electronic products. This has motivated a surge in research on conduction physics, particularly focusing on novel materials and interfaces as well as dimensional scaling below electron and phonon mean-free-paths (MFPs) [2], [3]. The semiconductor industry has, remarkably, also provided the tools that have accelerated this learning. Some of the most exciting heat conduction measurements leverage high-fidelity nanofabrication and, most recently, the most informative conduction simulations use high-performance parallel computing to reveal molecular and atomistic physics.

Therefore, the semiconductor industry has motivated, enabled, and generally catalyzed a renaissance in conduction research, focusing on nanoscales and advanced materials, as well as a dramatic surge in knowledge and learning in the field of heat transfer. The rate of discovery about nanoscale heat conduction is paralleled only by that which occurred during a golden age in the early and mid-20[th] century, when physicists and engineers pioneered conduction measurement techniques and leveraged new quantum and semi-classical theories to understand conduction in bulk media [4]–[8]. We are in the middle of a similar surge in conduction research today, truly a *heat conduction renaissance*. This paper summarizes a few of the most provocative and interesting discoveries, with an emphasis on physics learning and transformations in our understanding of how heat moves in solids.

**Phonon Subcontinuum Effects**

At the heart of interesting nanoscale conduction physics is the sub-continuum nature of thermal transport, which becomes apparent when device length scales shrink below the average MFPs of heat-carrying phonons (~10-1000 nm in silicon [9]). In this regime, strong diffuse boundary scattering of phonons reduces thermal conductivity ($\kappa$), thereby limiting heat dissipation. While the study of phonon boundary scattering dates back many decades [7], [10]–[12], the specific investigation of silicon nanostructures (e.g. thin-films and nanowires) goes back about twenty years [13]–[19]. Another type of subcontinuum effect is related to the suppression in the effective thermal conductivity when the dimensions of a hot-spot are comparable to the phonon MFPs [16], [20].

In the past few years, these ideas have motivated new experiments to map out the *spectrum* of phonon MFPs, i.e. quantify the relative contributions of phonons with different frequencies (and hence, MFPs) to the total thermal conductivity. These include measurements of frequency-dependent thermal conductivity [21]–[23], where a reduced $\kappa$ is observed at high-frequencies due to the exclusion of phonons with MFPs longer than the thermal penetration depth. In addition, optical measurements using finite laser spot-size [20], transient-grating pitch [24], and periodically-patterned heaters [25], in conjunction with inverse solution methods [26] have enabled reconstructions of phonon MFP spectra in a wide variety of materials [27], [28]. Looking ahead, as these techniques become more standardized, it is conceivable that they will become part of the thermal engineer's toolkit, in a way that will enable the discovery of new materials for energy harvesting and thermal management.

**Search for Ultra-High Thermal Conductivity Materials**

A topic that has received significant attention lately is the search for crystalline materials with ultra-high thermal conductivity. Research in this area received tremendous impetus following theoretical papers by Lindsay and Broido *et al.* [29], [30], that predicted high room-temperature $\kappa$ in boron arsenide (BAs), nearly ~2200 Wm$^{-1}$K$^{-1}$, on par with single-crystal diamond. Fundamentally, this high $\kappa$ is driven by (in addition to the light atoms and strong bonds) the large phonon band gap between acoustic and optical modes, which reduces the scattering phase space for thermal phonons, and close bunching of acoustic modes, which decreases the probability of high-frequency acoustic-acoustic phonon Umklapp scattering. Although this prediction was revised recently by Feng *et al.* [31] to include the effects of four-phonon scattering (which reduces the room temperature $\kappa$ to ~1400 Wm$^{-1}$K$^{-1}$, below that of single crystal diamond), early experimental efforts have shown promise. The first thermal conductivity measurements of BAs by Lv *et al.* [32] obtained values of ~200 Wm$^{-1}$K$^{-1}$. More recent experiments [33] on higher quality BAs crystals have reached values closer to ~350 Wm$^{-1}$K$^{-1}$, approaching the thermal conductivity of copper. Another closely related material is boron phosphide, in which Kang *et al.* [34] measured $\kappa$ up to ~460 Wm$^{-1}$K$^{-1}$. With rapid progress being made in improving material quality, concurrent with advances in thermal characterization techniques, this promises to be an exciting area of research.

**Influence of Defects and Microstructure**

While the search for diamond-like high $\kappa$ materials continues, it is increasingly becoming evident that defects play a very significant role in limiting thermal transport. In the specific example of diamond heat spreaders, while single crystal diamond can have a $\kappa$ as high as ~2000-3000 Wm$^{-1}$K$^{-1}$ (the upper limit is for isotopically enriched diamond with ~0.1 % $^{13}$C [35]) most realistic applications use processes such as chemical vapor deposition (CVD) to grow material directly in regions where it is needed. This material is of special interest for improved thermal management of GaN-based power amplifiers and other high-power photonic and electronic components, where it is used for dissipating heat in the "near junction region" (i.e. within 100 μm of the electronic junction). Due to the high density of grain boundaries (GBs), $\kappa$ within the first few μm near the interface can be almost two orders of magnitude lower than the single crystal. Over the past several years, many studies have examined this strong effect of microstructure on heat transport in thin-film diamond. Recently, using time-domain thermoreflectance (TDTR) measurements of ~1 μm thick suspended polycrystalline diamond membranes, Sood *et al.* [36] showed that $\kappa$ is inhomogeneous across the thickness of the film due to the evolution of grain size during the growth process. More interestingly, the columnar texture of grains leads to a significant anisotropy in the $\kappa$ tensor; for 0.5-1 μm thick films, the in-plane $\kappa$ can be ~3-5x lower than the cross-plane $\kappa$. This strong anisotropy can have serious implications for heat-spreading in the immediate vicinity of hot spots, as it would lead to a preferential extraction of heat-flux in the vertical direction at least within the first few μm.

Using a fractal-like geometric model for grain growth and coalescence, Sood *et al.* [36] predicted the evolution of the thermal conductivity tensor in polycrystalline diamond. By including the effects of diffuse phonon scattering at the disordered GBs, and the columnar orientation of grains, this model successfully predicted both the anisotropy and thickness-dependent inhomogeneity in thin-film diamond, and estimated the GB thermal resistance to be ~1-2 m$^2$KGW$^{-1}$, which corresponds to an equivalent thickness of few μm of single-crystal diamond. The strong resistive influence of GBs in diamond has also been observed experimentally by other researchers [37]–[41], going back to pioneering work by Graebner and coworkers [42], [43]. It is evident that to achieve the promise of "ultra-high $\kappa$" in any crystalline material, it will be crucial to understand the role played by defects and microstructure in limiting thermal transport.

**"Visualizing" Phonon-Defect Interactions**

Over the past several years, many studies have examined the connection between grain structure and thermal transport in polycrystalline materials. However, typical experimental approaches used large thermal probes that averaged over the impacts of thousands of GBs within a single measurement. As electronic device dimensions scale down to length scales comparable with grain sizes of heat-spreading substrates, it will be important to understand how microstructure affects local thermal transport, in the immediate vicinity of individual GBs.

To accomplish this, Sood *et al.* [44] recently developed a scanning-TDTR method to quantitatively map spatially-varying $\kappa$ near GBs in large-grained boron-doped CVD diamond. A key experimental challenge in this measurement was the precise location of GBs, and the co-location of TDTR measurements with images of the underlying microstructure. This was achieved using electron backscatter diffraction (EBSD) in a scanning electron microscope (SEM). Using this experimental platform, Sood *et al.* [44] observed a nearly two-fold reduction in the local thermal conductivity immediately adjacent to single GBs in boron-doped CVD-diamond. While the influence of near-boundary disorder (including dopants, dislocations, and extended defects) on local thermal transport would certainly depend on the specific details of the material, this work shows that in general, it may be important to consider heterogeneities in the local thermal environment due to variations in the microstructure of the heat-spreading layer. More broadly, uncovering the fundamental physics of phonon-defect interactions at the nanoscale will be key to enabling the rational design of materials for a wide range of applications in thermal management and energy conversion.

**Low-Dimensional Materials: Anisotropic Heat Spreaders**

Two-dimensional (2D) materials have reinvigorated the search for new solid-state heat spreaders. Owing to their unique crystal structure, van der Waals layered materials have highly anisotropic thermal transport properties, with the in-plane $\kappa$ being up to two orders of magnitude higher than the cross-plane $\kappa$. Measurements of the in-plane thermal conductivity of suspended monolayer graphene have shown promise, with numbers in the range ~2000-4000 Wm$^{-1}$K$^{-1}$ at room temperature (depending on isotope concentration and grain size), among the highest of any known material [45]–[48].

While the thermal conductivity of a graphene monolayer on substrate was found to be lower than the free-standing layer, at ~600 Wm$^{-1}$K$^{-1}$[49] it is still higher than most metals (for example, compared to copper at ~400 Wm$^{-1}$K$^{-1}$). We note that due to the long thermal phonon MFPs, in-plane $\kappa$ in graphene "ribbons" of finite width can be suppressed due to boundary scattering (to ~100 Wm$^{-1}$K$^{-1}$ for a 65 nm wide ribbon), as shown by Bae *et al.* [50]. An interesting demonstration of graphene's heat-spreading capabilities was made by Yan *et al.* [51] who showed that a few-layer "graphene quilt" enabled the reduction in operating temperature of a high-power GaN transistor by ~20°C. Although graphene shows significant promise, due to its high electrical conductivity it may have limited applicability as a heat-spreader in most electronic devices. Consequently, recent efforts have focused on hexagonal boron nitride (hBN) as an electrically-insulating alternative. Jo et al. [52] measured high in-plane $\kappa$ of ~250-360 Wm$^{-1}$K$^{-1}$ for suspended 5-11 layer hBN, approaching previously reported values for the bulk material [53], while Choi et al. [54] demonstrated large reductions in hot spot temperature of graphene devices using thin layers (35-80 nm) of hBN.

Looking ahead, in order to realize the full potential of 2D heat spreaders for realistic applications, it will be important to develop methods to synthesize or transfer the materials at low temperatures (preferably below 450°C for CMOS integration), while maintaining high crystal quality. Furthermore, it will be crucial to engineer improved thermal interfaces between the 2D layers and substrate. Given that this is nominally a weak, van der Waals bonded interface, there are potential trade-offs in choosing layered heat spreaders. For example, a higher in-plane thermal conductivity might come at the cost of a lower interfacial thermal conductance [55], [56].

**Semiconductor Heterostructures: Coherent Phonons**

The efficiency of several semiconductor optoelectronic devices is often limited by heat dissipation bottlenecks. For example, in Quantum Cascade Lasers (QCLs), large volumetric rates of heat generation (exceeding ~1 GWcm$^{-3}$) within the active core can cause significant temperature rise (> 100°C), eventually resulting in low wall-plug efficiencies (as low as 10 %). The gain media in typical quantum well lasers consist of thin-film semiconductor superlattices, made of epitaxially-grown materials such as GaAs and AlAs. To uncover the factors limiting heat transport within these devices, it is crucial to understand the fundamental processes governing phonon transport within thin films and across interfaces.

Cross-plane phonon transport in superlattices has been studied extensively by several groups over the past two decades [57]–[65]. Recently, there has been a resurgence of interest in this topic, specifically on the question of coherent phonons. In 2012, Luckyanova et al. [62] observed that the cross-plane $\kappa$ of epitaxially-grown GaAs/AlAs superlattices increases monotonically with the total film thickness (= period thickness × number of periods), implying that there are long MFP phonons that preserve phase and coherently transmit across multiple interfaces. This contrasts with what is typically expected in the case of incoherent transport, where the thermal conductivity is independent of the total sample thickness. A complementary approach to probing coherent transport is to examine the dependence of $\kappa$ on superlattice period thickness. While this dependence is strictly monotonic in the incoherent regime, Mahan et al. [66] predicted that in the coherent regime, wave-like interference of phonons would lead to a non-monotonic dependence on period thickness. This is a consequence of band-folding and the formation of mini-Umklapp Brillouin zones, which affects phonon scattering rates and group velocities. Ravichandran et al. [61] made an experimental demonstration of this effect, showing that the cross-plane $\kappa$ of high-quality perovskite oxides has a minimum at a period thickness of ~2 nm.

These results have generated much excitement in the heat transfer community, as they suggest potential opportunities to control heat flow using the wave-like nature of phonons [67], [68]. Looking ahead, a key question in the field of "phononics" is whether these wave-like effects can manifest at room temperature, where dominant heat-carrying acoustic modes have frequencies in the THz range, and consequently wavelengths that are smaller than a few nm. Recently, using measurements of thermal conductivity in suspended silicon nano-meshes with varying degrees of disorder, Lee et al. [69] and Maire et al. [70] showed the near-absence of coherent effects at temperatures above ~15 K. At $T < 15$ K, Maire et al. [70] measured a lower $\kappa$ in meshes with an ordered array of holes, indicating the existence of a phononic band gap due to wave interference. To utilize phonon wave effects for novel thermal management applications near room temperature will require the fabrication of structures with periodic features on the order of a few nm, with exceptionally smooth (i.e. specular) surfaces, and this remains an open challenge.

**Active Thermal Management: Heat Switches**

The ability to actively regulate heat flow at the nanoscale could have dramatic implications for thermal management of electronics [71]. Research in this area has seen a surge over the past ten years, concurrent with improvements in nano-thermal metrology techniques including *in situ* characterization capabilities. Broadly classified, these approaches have utilized temperature-induced phase-change [72], [73], electric-field induced modulations of domain-walls in ferroelectrics [74], magnetic-field induced re-orientation in liquid crystals, and electrochemical intercalation-induced modifications of material microstructure [75]–[77], to achieve reversible tuning of thermal conductivity. Typical "on/off" thermal conductance ratios are <2x, while switching speeds may vary depending on the method used (<100 ms for ferroelectric domain modulation, to hours for ion intercalation in a bulk crystal). We note that this list does not include microelectromechanical (MEMS) based switches which control heat flow by making and breaking mechanical contact. Arguably, the field of thermally functional materials is still in its early stages; however, it is becoming increasingly clear that for active regulators to have a significant impact, it will be important to increase both the thermal switching ratio and switching speed.

As a possible application, active thermal regulators can help to regulate the temperature fluctuations of an operating device by dynamically adjusting their thermal resistance in response to variations in heat flux. Reducing temperature fluctuations can have a dramatic impact on the lifetime of a device. To first order, a thermal transistor with an on/off ratio of $\gamma$ will reduce the amplitude of temperature oscillations $\Delta T$ by a factor $\sim 1/\gamma$. Given that the relationship between $\Delta T$ and the number of cycles to failure $N_f$ is highly nonlinear [78] (for example $N_f \sim \Delta T^{-3.5}$), it is evident that increasing the thermal on/off ratio can have a significant impact on device lifetime and reliability.

Looking ahead, the challenge is to engineer nanoscale devices with on/off ratios approaching at least one order of magnitude [79]. Such a breakthrough can have transformative applications and enable the control of heat flow using thermal circuits, analogous to electronic circuits. It is conceivable that such an advance will require a combination of two or more of the approaches noted above, for example, a coupling between electrochemical and mechanical degrees of freedom to achieve active thermal control.

**Compliant Heat Spreaders**

To address thermal management challenges at the packaging level, there is a need to engineer materials that are thermally conductive, yet mechanically compliant [80]. One approach towards this goal is to create composite materials that combine soft, low $\kappa$ materials with hard, high $\kappa$ fillers. However, significant enhancements using this approach are typically achieved only at high filler concentrations above the percolation threshold; this can make the composite electrically conducting and, in some cases, prohibitively expensive.

Another promising approach is to start with polymers that are natively mechanically compliant and engineer their intrinsic microstructure to increase their thermal conductivity. This could be done broadly through increasing chain alignment or by engineering stronger inter-chain interactions. Following the first approach, Kurabayashi *et al*. [81] observed highly anisotropic thermal conductivity in spin-coated polyimide films as thin as 500 nm and showed that the in-plane $\kappa$ can be as high as ~1.7 Wm$^{-1}$K$^{-1}$, nearly ~7x higher than the cross-plane value. In 2010, Shen *et al*. [82] demonstrated a breakthrough enhancement by stretching individual nanofibers of polyethylene, achieving metal-like $\kappa$ exceeding ~100 Wm$^{-1}$K$^{-1}$. Furthermore, their theoretical calculations showed that the fundamental limit of thermal conductivity in single polyethylene chains is ~180 Wm$^{-1}$K$^{-1}$, suggesting that there is scope for significant enhancements using this approach. Recently, progress has been made in developing a scalable platform to manufacture large sheets of highly-aligned polyethylene [83] and arrays of aligned electropolymerized polythiophene nanofibers [84], showing thermal conductivity enhancements of ~60x and ~20x, respectively, over the bulk polymer. Taking another approach, other groups have made progress towards engineering high $\kappa$ polymers by engineering inter-chain bonding interactions [85], [86], achieving thermal conductivity enhancements as high as ~20x. Overall, these studies have demonstrated the promise of molecularly engineered polymers as scalable, inexpensive, and mechanically robust thermal management solutions at the packaging level.

In summary, this paper has reviewed some of the key developments that have taken place in the field of nanoscale thermal conduction physics over the past two decades, highlighting in each case, inspirations drawn from the electronics and packaging communities. In doing so, we have also described some of the big open questions facing the thermal management community, which provide exciting opportunities for technological innovation and fundamental discovery.

**Acknowledgments**

The authors acknowledge the outstanding contributions of members of the Stanford NanoHeat group over the years, especially those that have contributed to ITherm in academic and organizational capacities. We also acknowledge numerous collaborators in the semiconductor industry, as well as the leadership and vision of scholars in the nanoscale heat transfer community generally. We acknowledge also the generous support and sponsorship of semiconductor companies, government laboratories, and national agencies over the past 25 years.